\documentclass[]{aa}
\usepackage[dvips]{graphicx}
\usepackage{natbib}
\bibpunct{(}{)}{;}{a}{}{,}
\usepackage{rotating}

\newcommand{\Msun}{M$_\odot$}
\newcommand{\Rsun}{R$_\odot$}
\newcommand{\Lsun}{L$_\odot$}

\newcommand{\Mbol}{$M_{\rm bol}$}
\newcommand{\kms}{km~s$^{-1}$}
\newcommand{\Ha}{H$_\alpha$}

\begin{document}

\title{The Henize sample of S stars}
\subtitle{IV. New symbiotic stars\thanks{Based on
    observations carried out at the European Southern Observatory
    (ESO, La Silla, Chile; program 60.E-0805) and at the Swiss 70 cm telescope
    (La Silla, Chile)
    }
}
\author{
S. Van Eck\thanks{Post-doctoral Researcher,
F.N.R.S., Belgium}
\and
A. Jorissen\thanks{Research Associate,
F.N.R.S., Belgium}\\
}
\offprints{S. Van Eck; email: svaneck@astro.ulb.ac.be}
\institute{
Institut d'Astronomie et d'Astrophysique, Universit\'e Libre de
Bruxelles, CP 226, Boulevard du Triomphe, B-1050 Bruxelles,
Belgium
}

\date{Received / Accepted }

\abstract{
The properties of the few symbiotic stars detected among the 66 binary S
stars from the Henize sample are discussed. Two stars (Hen 18 and Hen
121) exhibit both a strong blue-violet continuum and strong \Ha\
emission (FWHM of 70 \kms), whereas Hen 134 and 137 exhibit
weak \Ha\ emission. The \Ha\ profiles are typical of
non-dusty symbiotic stars belonging to class S-3 as defined by Van Winckel
et al. (1993, A\&AS 102, 401). In that class as in the Henize symbiotic S
stars,  He I, [N II] or [S II] emission lines are absent, suggesting that the
nebular density is high but the excitation rather low. The radial
velocity of the centre of the \Ha\ emission is identical to that of
the companion star (at least for Hen~121 where  this can be checked
from the available orbital elements), thus  suggesting that the 
\Ha\ emission  originates from gas moving with the companion star. 
For Hen~121, this is further confirmed by the disappearance of the
ultraviolet Balmer continuum when the companion is eclipsed by the S star. 
Hen~121 is thus the second eclipsing binary star discovered among
extrinsic S stars (the first one is HD~35155). 
A comparison of the available data on orbital periods and \Ha\
emission leads to the conclusion that \Ha\ emission in S stars seems to
be restricted to binary systems with periods in the range 600 --
1000~d, in agreement with the situation prevailing for red symbiotic stars
(excluding symbiotic novae).   
Symbiotic S stars are found among the most evolved extrinsic
S stars.

}

\titlerunning{New symbiotics among S stars}
\maketitle

\section{Introduction}
\label{Sect:intro}
The defining spectral properties of symbiotic and S stars are very 
different, and do not {\it a priori} hint at a possible relationship 
between them. Symbiotic stars exhibit at the same time
spectral features typical 
of cool stars (like molecular bands), of hot stars (like an ultraviolet
continuum) and of nebulae (like forbidden emission lines) 
\citep[see][]{Kenyon-86}.  
Stars of type S are red giants with ZrO bands in their spectra. 
However, technetium-poor\footnote{Tc-rich (also called `intrinsic') 
S stars are genuine asymptotic giant branch stars
  that, unlike Tc-poor S stars, are not necessarily binaries, and are
  therefore not relevant to the present
study} (or `extrinsic')  S stars 
\citep[see][for recent reviews]{Jorissen-VE-98,VanEck-Jorissen-99c}  and symbiotic stars 
 share similarities in that they 
are two classes of binary stars involving a red
giant and a white dwarf companion in relatively wide systems with
orbital periods larger than about 200~d.  This similarity naturally
raises the question of the relationship between these two families, or in
other words, why do those binary systems seem to appear in two different
flavours? This question is actually twofold: (i) Do symbiotic systems
exhibit the same kind of abundance peculiarities as S stars, namely
enhanced carbon and  elements heavier than Fe produced by the s-process of
nucleosynthesis? (ii)  Do Tc-poor S stars exhibit symbiotic activity?

To answer the first question, it is necessary to distinguish yellow
symbiotics (i.e., involving a G or K giant) from red symbiotics (i.e., 
involving an M giant). All detailed abundance studies performed so far 
for yellow symbiotics have revealed heavy-element overabundances
resembling those observed in S stars \citep{Schmid-94, Smith-96,
Smith-97, Pereira-97, Pereira-98, Pereira-01, Pereira03}. 
On the contrary, not a single red symbiotic star from the catalogue
of \citet{Muerset-99} seems to exhibit the ZrO distinctive spectral
features of S-type stars. This puzzling situation is reviewed by \citet{Jorissen03}.

The present paper deals with the question of the possible symbiotic
activity among S stars. Several previous studies have indeed
identified S stars with symbiotic activity (as will be reviewed in 
Sect.~\ref{Sect:discussion}), but those previously known cases were
isolated cases, not resulting from a systematic search. Hence they do
not provide much insight into the frequency of this phenomenon among
S stars. In this paper, we report the result of a search for \Ha\ emission
among a sample of 29 supposedly binary S stars identified by 
\citet{VanEck-Jorissen-99b,VanEck-Jorissen-99c} (hereafter denoted by
Papers~II and III, respectively) in
the Henize sample of 205 S stars.

\section{Binary S stars among the Henize sample}
\label{Sect:Henize}

The Henize sample of S stars has been assembled by K. Henize from
the objective-prism plates of the Michigan-Mount Wilson survey of the
southern sky for \Ha\ emission objects. Besides this primary aim, 
Henize readily realized that the plates could also be used 
to identify S stars (not necessarily with \Ha\ in emission). His list
of  S stars, that he never published however,
can be traced through Stephenson's General Catalogue of Galactic S
Stars \citep{Stephenson-84}, where they appear as {\it Henun nnn}\footnote{In
SIMBAD, they are referred to as {\it Hen 4-nnn}} (with $nnn \le 205$).
The major advantage of the Henize sample is that it is not restricted
to the galactic plane, since it comprises the 205 S stars south of
declination $-25^\circ$ that are brighter than $R = 10.5$ (the
completeness of the sample has been discussed in Paper~III).
Therefore, the sample is {\it not} biased against extrinsic,
Tc-poor, binary  S stars, which are not as concentrated along the galactic
plane as are intrinsic S stars (Paper III).
The identification of the extrinsic S stars in the Henize sample has
been performed in Papers~II and III using multivariate classification
analysis relying on the radial-velocity dispersion, the $U-B$, $B-V$,
$V-K$, and IRAS $K-[12]$ color indices, as well as on the ZrO and TiO band
strengths, on the presence/absence of Tc and on the photometric variability.
This classification procedure yielded 66 extrinsic S stars out of 199
S stars with available data (see Table~6 of Paper~II). The binary
nature of those stars is strongly suggested by the large dispersion of 
their radial velocities (Fig.~10 of Paper~III), while orbits are
available for 5 of those stars (Table~1 of Paper~II).

\begin{figure}[]
\resizebox{\hsize}{!}{\includegraphics{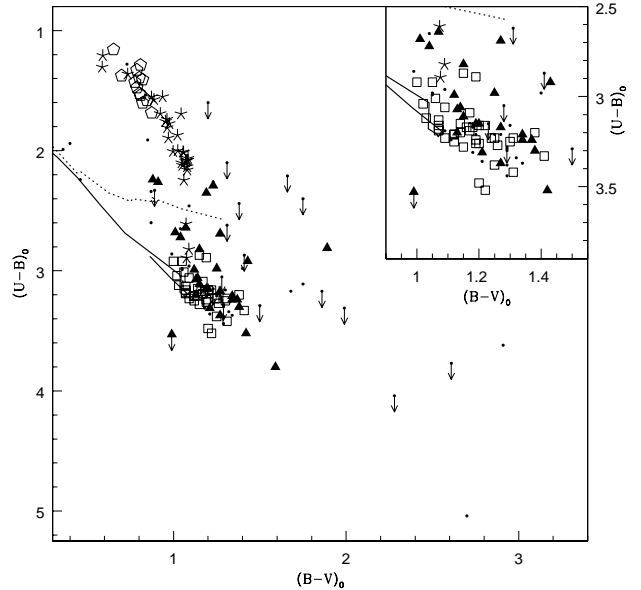}}
\caption{\label{Fig:photsymbio} 
Dereddened $(U-B,B-V)$ diagram,
in the Geneva photometric system, for 
non-symbiotic S stars (average of individual measurements)
and symbiotic S stars (individual measurements).
Tc-rich, intrinsic S stars and Tc-poor, extrinsic (binary) S stars are
represented by  
filled triangles and open squares, respectively,
whereas dots denote stars with unknown Tc content. 
Stars with only upper limits on their $U$ flux
are flagged with an arrow.
The individual measurements of the symbiotic S stars Hen~18 and~121
are represented by star symbols and open hexagons, respectively.
Note how the color sequence of Hen~121 extends down into the region
occupied by the extrinsic S stars. The three
measurements of Hen~121 located in the inset were taken during the
eclipse of the companion by the S star.   
The solid and dotted lines are the normal giant and dwarf sequences, 
   respectively, from \protect\cite*{Grenon-78}
}
\end{figure}

\renewcommand{\baselinestretch}{1}
\begin{table*}
\caption{\label{Tab:photsymbio}
Geneva photometry of Hen 18 and Hen 121. 
$IPV$ and $IPC$ are quality indices for the $VM$ magnitude and for the
colors, respectively, in a scale from 0
(bad) to 4 (excellent). Data in bold case correspond 
to the eclipse of the Balmer ultraviolet continuum of Hen~121
}
\begin{tabular}{rrrrrrrrrrr}
\hline
\medskip\\
\multicolumn{1}{c}{JD}     &    IPV     &    $VM$ & IPC  &     $U-B$ &     $V-B$ &   $B1-B$ &   $B2-B$ &   $V1-B$  &   $G-B$  & airmass\cr
(-2\ts440\ts000) \cr
\hline
\medskip\\
\noalign{Hen 18}
\medskip\\
 8686.573    &     3  &   10.488  &  3  &    1.288 & -0.852 &  1.196  & 1.208 & -0.041 &  0.085  &1.02 \cr   
 8718.558    &     3  &   10.590  &  3  &    1.559 & -0.976 &  1.297  & 1.177 & -0.167 & -0.045  &1.21 \cr   
 9428.543    &     2  &   10.508  &  3  &    1.668 & -1.010 &  1.331  & 1.141 & -0.195 & -0.078  &1.02 \cr   
10148.577    &     3  &   10.605  &  3  &    1.814 & -1.073 &  1.413  & 1.158 & -0.249 & -0.143  &1.03 \cr   
10166.532    &     3  &   10.603  &  3  &    1.653 & -1.003 &  1.345  & 1.186 & -0.189 & -0.071  &1.03 \cr   
10170.535    &     2  &   10.571  &  2  &    1.710 & -1.046 &  1.373  & 1.170 & -0.232 & -0.108  &1.06 \cr    
10181.520    &     3  &   10.602  &  4  &    1.604 & -0.987 &  1.291  & 1.153 & -0.176 & -0.045  &1.10 \cr    
10189.507    &     2  &   10.603  &  3  &    1.538 & -1.020 &  1.291  & 1.136 & -0.211 & -0.082  &1.12 \cr    
10506.607    &     2  &   10.743  &  2  &    1.505 & -0.899 &  1.278  & 1.187 & -0.082 &  0.066  &1.04 \cr   
10727.839    &     4  &   10.697  &  4  &    1.731 & -1.021 &  1.320  & 1.168 & -0.213 & -0.096  &1.20 \cr    
10741.800    &     4  &   10.625  &  4  &    1.457 & -0.987 &  1.299  & 1.168 & -0.171 & -0.050  &1.21 \cr    
10788.770    &     2  &   10.563  &  3  &    1.523 & -1.002 &  1.342  & 1.144 & -0.194 & -0.070  &1.00 \cr    
10798.752    &     2  &   10.524  &  3  &    1.417 & -1.008 &  1.283  & 1.118 & -0.183 & -0.069  &1.00 \cr
\medskip\cr    
\noalign{Hen 121}
\medskip\\
 8685.809    &     3  &   10.400 &   3  &    1.809  &-1.034 &  1.226 &  1.195 & -0.222 & -0.111 & 1.03    \cr 
 8715.665    &     2  &   10.551 &   2  &    1.602  &-1.017 &  1.204 &  1.262 & -0.194 & -0.081 & 1.01    \cr 
 9044.830    &     2  &   10.626 &   2  &    1.627  &-0.967 &  1.199 &  1.243 & -0.156 &  0.019 & 1.03    \cr 
{\bf 9053.761} &     3  &   10.384 &   3  &{\bf 2.663}&-1.153 &  1.249 &  1.158 & -0.334 & -0.178 & 1.00    \cr 
{\bf 9068.748} &     3  &   10.468 &   3  &{\bf 2.873}&-1.169 &  1.361 &  1.175 & -0.357 & -0.209 & 1.01    \cr 
{\bf 9074.728} &     2  &   10.509 &   2  &{\bf 2.945}&-1.156 &  1.344 &  1.185 & -0.344 & -0.183 & 1.01    \cr 
10145.706    &     2  &   10.454 &   1  &    2.059  &-1.077 &  1.255 &  1.187 & -0.274 & -0.134 & 1.06    \cr 
10150.830    &     3  &   10.443 &   2  &    1.861  &-1.044 &  1.280 &  1.212 & -0.234 & -0.108 & 1.09    \cr 
10154.780    &     1  &   10.437 &   3  &    1.826  &-1.055 &  1.254  & 1.207 & -0.243 & -0.106 & 1.02    \cr 
10163.653    &     2  &   10.551 &   1  &    1.792  &-1.042 &  1.246 &  1.235 & -0.229 & -0.094 & 1.08    \cr 
10166.679    &     3  &   10.563 &   3  &    1.744  &-1.003 &  1.219 &  1.205 & -0.187 & -0.042 & 1.02    \cr 
10170.761    &     2  &   10.576 &   2  &    1.606  &-0.952 &  1.184 &  1.248 & -0.126 &  0.006 & 1.05    \cr 
10179.664    &     3  &   10.604 &   3  &    1.946  &-1.050 &  1.218 &  1.242 & -0.223 & -0.085 & 1.00    \cr 
10187.636    &     3  &   10.442 &   2  &    2.173  &-1.132 &  1.269 &  1.183 & -0.334 & -0.195 & 1.01    \cr 
10212.517    &     3  &   10.526 &   3  &    2.213  &-1.159  & 1.289 &  1.186 & -0.344 & -0.213 & 1.08    \cr 
10225.598    &     2  &   10.672 &   3  &    1.923  &-1.103  & 1.227 &  1.237 & -0.261 & -0.171 & 1.04    \cr 
10255.522    &     3  &   10.429 &   4  &    2.117  &-1.164 &  1.307 &  1.154 & -0.359 & -0.236 & 1.05    \cr 
10257.520    &     3  &   10.414 &   3  &    2.133  &-1.163 &  1.275 &  1.162 & -0.340 & -0.220 & 1.06    \cr 
10259.500    &     3  &   10.420 &   3  &    2.133  &-1.154 &  1.275 &  1.155 & -0.344 & -0.219 & 1.03    \cr 
10262.531    &     3  &   10.446 &   4  &    2.299  &-1.140 &  1.365 &  1.186 & -0.315 & -0.202 & 1.12    \cr 
10268.517    &     2  &   10.391 &   2  &    2.062  &-1.135 &  1.303 &  1.152 & -0.329 & -0.180 & 1.13    \cr 
10272.492    &     3  &   10.414 &   3  &    2.052  &-1.104 &  1.303 &  1.185 & -0.298 & -0.142 & 1.09    \cr 
10273.530    &     3  &   10.418 &   3  &    2.153  &-1.127 &  1.270 &  1.193 & -0.321 & -0.163 & 1.24    \cr 
10274.521    &     4  &   10.428 &   4  &    2.141  &-1.150 &  1.293 &  1.143 & -0.333 & -0.195 & 1.20    \cr 
10285.492    &     3  &   10.477 &   3  &    2.069  &-1.138 &  1.252 &  1.198 & -0.331 & -0.192 & 1.21    \cr 
10288.498    &     3  &   10.471 &   2  &    2.186  &-1.159 &  1.295 &  1.187 & -0.342 & -0.200 & 1.29    \cr 
10291.518    &     3  &   10.482 &   3  &    1.746  &-1.123 &  1.271 &  1.211 & -0.298 & -0.184 & 1.49    \cr 
10480.809    &     1  &   10.239 &   1  &    1.415  &-0.814 &  1.197 &  1.257 & -0.018 &  0.152 & 1.03    \cr 
10514.882    &     3  &   10.283 &   3  &    1.260  &-0.670 &  1.121 &  1.279 & 0.128  & 0.298  &1.27    \cr 
10518.799    &     4  &   10.230 &   4  &    1.357  &-0.667 &  1.104 &  1.288 &  0.133 &  0.306 & 1.04    \cr 
\hline
\end{tabular}
\end{table*}

Among these 66 binary stars, two (namely Hen 18 and Hen 121) stand out 
because of their very blue colors (see Sect.~3.2 and Fig.~3 of
Paper~III). Fig.~\ref{Fig:photsymbio} presents the evolution of the  
color indices of these two stars in the dereddened\footnote{the
dereddening procedure applied to the data of Table~\ref{Tab:photsymbio}
to obtain Fig.~\ref{Fig:photsymbio} has been described in Sect.~3.3 of
Paper~II} $(U-B, B-V)$ diagram.
This evolution is very similar to the one of classical symbiotic stars
\citep[see e.g.,][]{Arkhipova-Noskova-85}.   
Table~\ref{Tab:photsymbio} lists the individual photometric
measurements in the Geneva system, which are too sparse 
to derive a meaningful lightcurve.
For Hen~121, our photometric data nevertheless contain the clear
  signature  of an eclipse of the ultraviolet Balmer continuum by the
  S star. This is clearly visible in Table~\ref{Tab:photsymbio}
  (numbers in bold face) and in the inset of
  Fig.~\ref{Fig:photsymbio}, which reveal that the $U-B$ color of
  Hen~121 becomes much redder at JD
  = 2\ts449\ts053.76, 2\ts449\ts068.75 and 2\ts449\ts074.73, and is then
  typical of non-symbiotic, extrinsic S stars. 
  The orbital elements listed in Paper~II indicate that these
  measurements were taken at phases 0.70, 0.72 and 0.73 ($\pm 0.01$),
  while the spectroscopic ephemeris predicts the eclipse of the
  companion by the S star to be central at phase 0.75. 
  This provides a clear indication that the Balmer continuum observed
  in the Hen~121 system is tied to the companion. 
  Hen~121 is thus the second eclipsing binary star discovered among
  extrinsic S stars \citep[the first one is HD~35155;][]{Jorissen-92c}.

The time of first contact (which occurred between JD= 2\,449\,044.83
and JD=2\,449\,053.76 according to Table~\ref{Tab:photsymbio} and 
Fig.~\ref{Fig:photsymbio}, corresponding to phases 0.69 and 0.70,
respectively) makes it possible to estimate the S-star radius.  
Adopting $\phi = 0.695$ for the phase of first contact, 
$\sin i = 1$, 
$M_{\rm S}=1$~\Msun\ and $M_{\rm WD}=0.6$~\Msun\ for the S-star and WD 
masses, respectively (as suggested by
Fig.~\ref{Fig:Hafit} below and consistent with the mass function 0.092 
M$_\odot$ for the system; see Paper~I), the radius of the S star
amounts to $R=153$~\Rsun, and its Roche lobe to 180~\Rsun.
Hence Hen~121 is close to filling its Roche lobe.

This radius, combined with an effective temperature of $3400\pm100$~K,
as derived from the [$T_{\rm eff}, (V-K)_0$] calibration of
\citet{Plez-03} with $(V-K)_0 = 6.1$,
yields \Mbol $=-3.9$. This combination of bolometric magnitude and
effective temperature locates Hen~121 among 
the most evolved extrinsic S stars, according to Fig.~2 of
\citet{VanEck-1998}. The mass loss rate of Hen~121 may concomittantly
be expected to be large, a fundamental requirement to trigger
symbiotic activity as discussed in Sect.~\ref{Sect:discussion}.

The high-resolution \Ha\ spectroscopic survey of extrinsic 
S stars presented in this paper confirms that the two stars standing
out in the $(U-B, B-V)$ diagram of Fig.~\ref{Fig:photsymbio}, namely
Hen~18 and Hen~121, are the
only two stars 
exhibiting clear symbiotic activity (two marginal cases -- Hen 134 and 
Hen 137 -- will be discussed in
Sect.~\ref{Sect:analysis}).
In fact,  when the multivariate classification analysis described
in Paper~III is allowed to go beyond the simple
extrinsic/intrinsic dichotomy, the extrinsic sample is further
subdivided in three subgroups, one of these (cluster \# 3 in Table~2
of Paper~III) comprising the two symbiotic S stars with
exceptionally blue and variable colors (see also Fig.~\ref{Fig:VK} below).   

\section{Observations}

High-resolution spectra (with $R = \lambda/\Delta\lambda = 80\ts000$)
were obtained between February 21 and 27, 1998, with the
1.4m~CAT telescope of the European Southern Observatory (ESO, La
Silla, Chile), equipped with the Coud\'e Echelle Spectrograph (CES)
and the $f/4.7$ long camera. The CCD detector (ESO \# 38) is a 
Loral/Lesser back-illuminated, UV flooded thin chip with 2688 $\times$
512 pixels of 15 by 15 $\mu$m.   
Spectra covering about 6~nm around the \Ha\ line ($\lambda$ 657.3~nm)
have been obtained for 29 extrinsic S stars. For the two symbiotic
candidates (Hen 18 and 121), spectra were also obtained at the central 
wavelength $\lambda$669.7~nm, covering  [N II] $\lambda 654.8$ and $\lambda
658.3$, 
\ion{He}{I} $\lambda 667.8$, and [S II] $\lambda 671.6$ and $\lambda 673.1$.
Spectra with these two settings were also obtained for the S stars
HD~35155 and HD~49368 that were previously reported to exhibit some
symbiotic activity (see Sect.~\ref{Sect:discussion}).  
At least two radial-velocity standard stars were observed each night
to tie our wavelength scale to the IAU one. 

\section{Analysis}
\label{Sect:analysis}

\subsection{\Ha\ profiles}

\begin{figure}[]
\resizebox{\hsize}{!}{\includegraphics{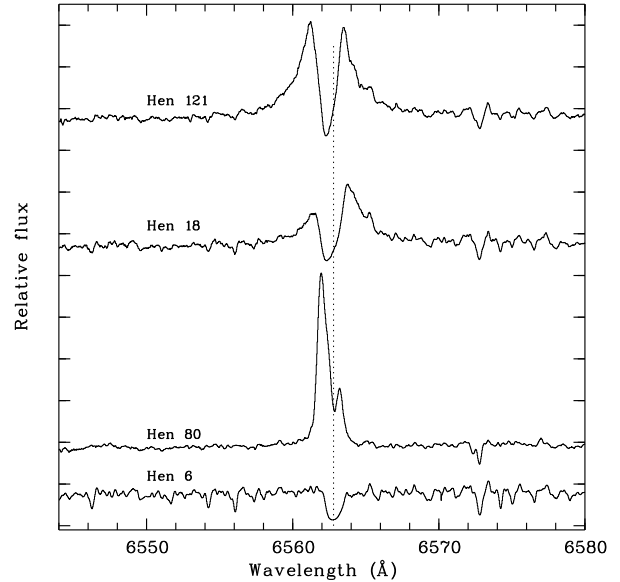}}
\vspace{-4cm}
\caption{\label{Fig:HaHen18-121}
The \protect\Ha\ profiles of the two extrinsic S stars Hen 18 and 121, 
taken on February 21, 1998,
compared to
typical \protect\Ha\ profiles for the intrinsic Mira S star Hen~80 and 
for the extrinsic S star Hen~6. The wavelength scale  has been corrected from the
radial-velocity of the stars, with the dashed line corresponding to the \Ha\ laboratory
wavelength }
\end{figure}

\begin{figure}[]
\resizebox{\hsize}{!}{\rotatebox{-90}{\includegraphics{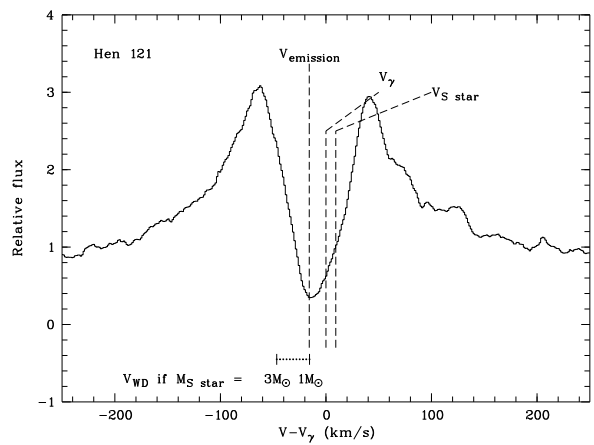}}}
\caption{\label{Fig:Hafit}
The velocities of the center of mass of the system (denoted $V_\gamma$), of the
S star (denoted $V_{\rm S star}$) and of the companion star (as indicated by
the horizontal segment corresponding to different values for $M_{\rm S}$, assuming $M_{\rm WD}
= 0.6$~\Msun) have been located on the \Ha\ profile of 
Hen 121. The central wavelength of the emission (dashed line labeled
$V_{\rm emission}$, as derived from a
gaussian fit to the emission profile) is seen to coincide with the velocity of the
companion, provided that $M_{\rm S} = 1$~\Msun\ and $M_{\rm WD}
= 0.6$~\Msun. The components' velocities have been computed from the
orbital  elements provided in Paper~II, given the orbital phase of  0.08, {\it i.e.,} close to
quadrature }
\end{figure}

Strong and broad \Ha\ emission lines were only observed for the two
extrinsic S stars already suspected to be symbiotics from their $UBV$ color
indices (Sect.~ \ref{Sect:Henize} and Fig.~\ref{Fig:photsymbio}). 
The corresponding profiles,
compared to those of a typical Mira, intrinsic S star (Hen 80) and of
a typical extrinsic S star (Hen 6), are shown
in Fig.~\ref{Fig:HaHen18-121}. The Hen~18 and Hen~121 profiles are
indeed very different from those of Mira stars [see also
\cite{Woodsworth-95} for a discussion of \Ha\ emission  
profiles of Mira S stars], both in terms of the
relative strengths of the red and blue emission peaks, and in terms of
the base width. 
The Hen~18 and 121 profiles consist of  a broad emission 
(half width at the base of about 250 \kms, and FWHM of 70 \kms) with a central
absorption core exactly centered on the emission (Fig.~\ref{Fig:Hafit}).  Those \Ha\
profiles are similar to the S-3 types identified by
\cite{VanWinckel-93} in symbiotic stars.
No He~I, [N II] or [S II]  emission lines are recorded in Hen~18 and
121, as expected for symbiotic stars with \Ha\ profiles of the S-3
kind [Table~6 of \cite{VanWinckel-93}]. The absence of these lines suggests that the
associated nebula has a high density and a rather low excitation state. The S-3 class
contains mainly  recurrent and symbiotic novae.

\begin{figure}[]
\resizebox{\hsize}{!}{\rotatebox{-90}{\includegraphics{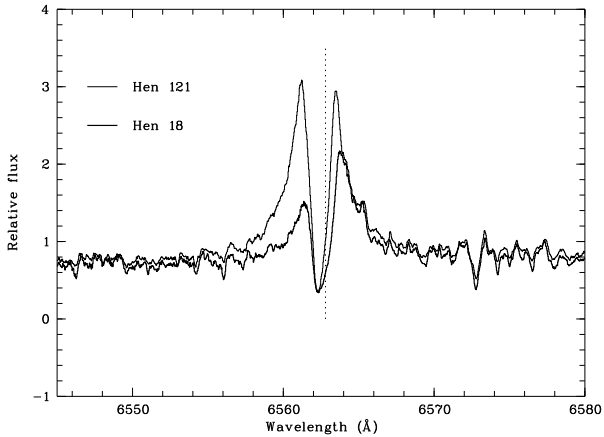}}}
\caption{\label{Fig:zoomHen18Hen121}
Comparison of the \protect\Ha\ profiles of Hen~18 and Hen~121
}
\end{figure}

Fig.~\ref{Fig:zoomHen18Hen121} compares the \Ha\ emission profiles of Hen~18 and
Hen~121, and reveals striking similarities. The slight difference between these
two profiles is reminiscent of the variations observed with orbital phase in symbiotic
systems 
\citep[for one particular example, see Fig.~4 of][]{Muerset-00}. Thus,  
similar formation mechanisms for \Ha\ must be operating in classical symbiotic stars and in  
symbiotic S stars. This mechanism is still debated. Two different
processes have been proposed: (i) the whole \Ha\ profile is formed in the accretion disc
forming around the accreting companion \citep[see e.g.,][]{Robinson-94}, (ii) the central
dip of the
\Ha\ profile results from absorption by the red-giant wind 
\citep{Schwank-97,Muerset-00} while the broad \Ha\ wings are caused by Raman
scattering of Ly$_\beta$ photons from the hot component \citep{Lee00}.

The data available for Hen 18 and Hen 121 are too scarce to distinguish between
these two formation processes, especially since the variation of the profile with
orbital phase is not known. Let us just remark that, in Hen~121,   
the broad emission is almost exactly 
centered on the velocity of the companion star (see Fig.~\ref{Fig:Hafit}), provided that
$M_{\rm S} = 1$~\Msun\ and $M_{\rm WD} = 0.6$~\Msun. This pair is
  consistent with the mass function of the system ($0.092\pm0.013$~M$_\odot$; 
see Table~1 of Paper~II), which predicts $M_{\rm S} =
0.93\pm0.1$~M$_\odot$ for an eclipsing system with 
$\sin i \sim 1$ and $M_{\rm WD} = 0.6$~M$_\odot$.
Moreover, despite its
blueshifted appearance, the absorption core  is exactly centered on the emission (the
false impression that it is blueshifted is actually caused by  the presence  of
\Ha\ and TiO absorptions formed in the atmosphere of the S star and which mutilate
the red wing of the \Ha\ emission). These properties of the \Ha\ profile tend to support
the hypothesis that the \Ha\ emission 
originates from gas close to the companion.  However, the
similarity between the velocity of the emission and absorption components in Hen~18 and
Hen~121, as revealed by Fig.~\ref{Fig:zoomHen18Hen121}, is difficult to understand in that
framework, since it is very unlikely that the orbital phases at the time of observation -- and
hence the velocity of the companion star -- were identical in these two systems (this can
unfortunately not be checked as orbital elements are not available for Hen~18). 

\subsection{More symbiotic stars among Henize S stars}

\begin{figure}[]
\resizebox{\hsize}{!}{\includegraphics[260,60][570,712]{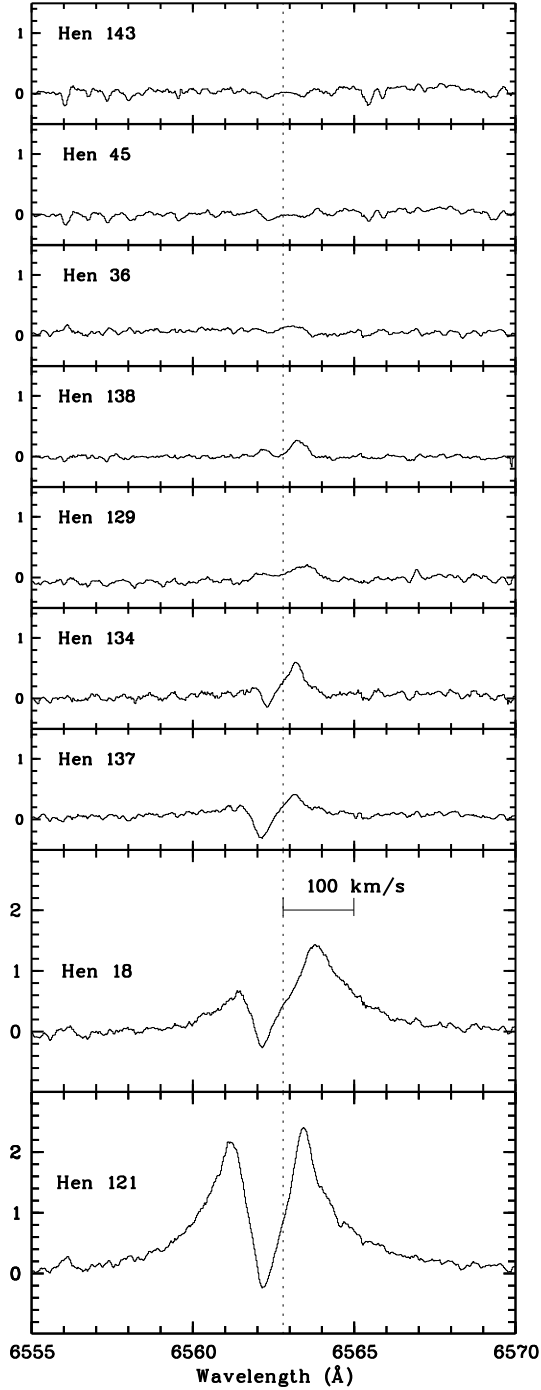}}
\caption[]{\label{Fig:HaHen6}
Residual \Ha\ profiles obtained  after subtracting the Hen~6 profile, for
a few representative cases (from top to bottom, ordered according to the increasing
strength  of their \protect\Ha\ emission): (i) extrinsic star with no residual \Ha\
emission/absorption (Hen 143); (ii) intrinsic stars where residual \Ha\
emission/absorption is not -- and must not be -- present (Hen  36 and 45); 
(iii) extrinsic stars with weak residual 
emission likely due to a slight mismatch
between the template spectrum (Hen~6) and the target spectra
(Hen~138 and 129; see Fig.~\ref{Fig:HaP} for more profiles of extrinsic stars); 
(iv) extrinsic
stars with weak symbiotic-like profiles (Hen 134 and 137); 
(v) symbiotic S stars (Hen 18 and 121).
The \Ha\ rest wavelength is indicated by a dashed line
}
\end{figure}

\begin{figure}[]
\resizebox{\hsize}{!}{\includegraphics[28,28][540,540]{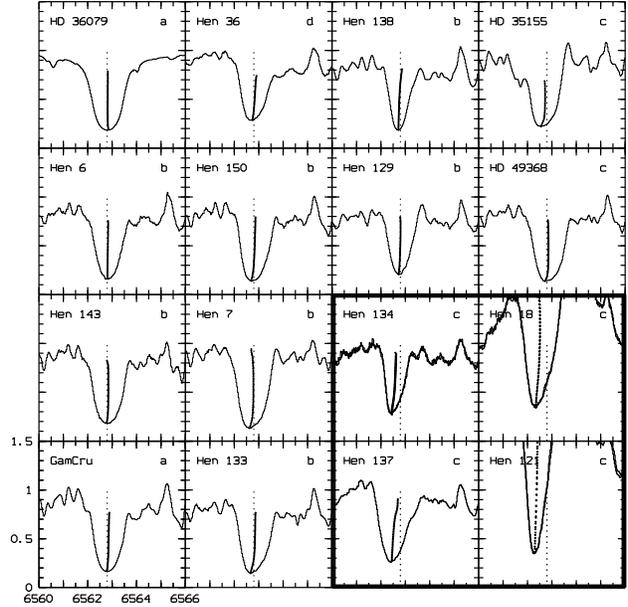}}
\caption{\label{Fig:bisector}
Representative examples of  \protect\Ha\ profiles and their associated
bisectors. 
All panels are drawn on the same scale, as indicated on the
lower left panel. All spectra have been corrected from the Doppler
shift, and the
\protect\Ha\ laboratory wavelength is depicted by the vertical dashed line.
The label in the upper right corner of each panel classifies the star 
according to the
following categories:  a: radial-velocity standard star; 
b: extrinsic S star;
c: extrinsic symbiotic S star (see also Table~\ref{Tab:activityP});
d: intrinsic S star.
The leftmost column corresponds to stars with an almost vertical
bisector; the second
to leftmost column displays stars with a bisector slightly slanted under 
the influence of
the TiO absorptions falling in the red wing of the \protect\Ha\ line; 
stars with a 
definite symbiotic signature in their \Ha\ profile, as discovered in the present study, 
lie within the bold frame
}
\end{figure}

In an attempt to find weaker cases of symbiotic activity among binary S stars from the
Henize sample, their \Ha\ profiles have been scrutinized 
in two different ways: (i) by deriving the residual profile after subtraction 
of the \Ha\ absorption normally present in S stars, and (ii) by    
examining the line bisector.

Fig.~\ref{Fig:HaHen6} presents a few representative examples of the
residual \Ha\ profile obtained after subtracting  the spectrum of
Hen~6, chosen as a template because its bisector appears to be very clean 
(see Fig.~\ref{Fig:bisector} below). Two new cases of S stars with symbiotic-like \Ha\
profiles emerge from this figure: Hen~134 and 137. 
The case of Hen~129 and Hen~138, also shown in Fig.~\ref{Fig:HaHen6},
is less clear: 
the weak emission remaining after subtraction of
the Hen~6 template (Fig.~\ref{Fig:HaHen6}) 
is more likely due to a template mismatch than to a real \Ha\ emission.

The P-Cygni shape of the residual 
\Ha\ profile appearing on Fig.~\ref{Fig:HaHen6} for Hen~134 and 137 may also be inferred
from their line bisectors displayed on  Fig.~\ref{Fig:bisector}, which are   
quite different from those of non-symbiotic stars.
Almost vertical, straight bisectors are found for the G giant HD 36079 and
the early M giant $\gamma$~Cru, as well as for the extrinsic S star Hen 6 (leftmost column of
Fig.~\ref{Fig:bisector}). 
TiO lines mutilating the red wing of the \Ha\ line  -- as confirmed by synthetic spectra -- are
responsible for the slanted appearance of the bisectors in the second to leftmost column in
Fig.~\ref{Fig:bisector}. This column includes intrinsic S stars and cool extrinsic S stars where TiO bands are strong.
In all these cases, the bisector in the core of the line is
nevertheless located very close to the 
\Ha\ wavelength, as expected. 
On the contrary, the bisectors of Hen~18 and 121 are strongly
blueshifted in the line core.
A similar trend is also clearly apparent in the case of Hen~134 and 137,
which we therefore tag as weak symbiotic stars.
All S stars for which our  \Ha\ profiles are indicative of symbiotic
activity have been collected in the four bold frames of Fig.~\ref{Fig:bisector}.
The analysis of the normalized \Ha\ profiles (Fig.~\ref{Fig:HaHen6})
and of the line bisector (Fig.~\ref{Fig:bisector}) thus yields 
the same set of symbiotic S stars among
the Henize sample. 

\subsection{Correlation of symbiotic activity with orbital period}
\label{Sect:discussion}

\begin{table*}
\caption[]{\label{Tab:activityP}
Presence or absence of symbiotic activity in S stars as a function of orbital period,
according to various diagnostics:  X-rays (`X'), UV
continuum  (in column `UV cont.', `y'
means that excess UV flux is present, but does not match a clean WD
spectrum), C~IV
$\lambda$155~nm and Mg~II $\lambda$280~nm lines, 
\protect\Ha\ emission, and He~I $\lambda$1083~nm.
The stars have been ordered according to increasing orbital period.
}
\vskip 0.5cm
\setlength{\tabcolsep}{3pt}

\begin{tabular}{lrcccccc}\hline
Name & $P$ & X
&UV & CIV & MgII & H$_\alpha$& He I\cr
& (d) & & cont. & $\lambda$ 155 & $\lambda$ 280 & em. &$\lambda 1083$
\cr
\hline
\cr
Hen 108 & 197 & ?  & ? & ?  & ?         & no & ?\cr
Hen 147 & 335 & ?  & ? & ?  & ?         &  no & ?\cr
HR 1105 & 596 & ?  & y & y & y  & ? & var\cr
Hen 137 & 636 & ?  & ? & ? & ?          & moderate & ? \cr
HD 35155 & 642 & var?& y & y & y  & moderate? & var\cr
Hen 121 & 764 & ?  & y & ?  & ? & strong& ?  \cr
HD 191226& 1210 & ?  & WD & no &  y & ? & wk em. \cr
Hen 119 & 1300 & ? & ? & ? & ? & no & ? \cr
HD 49368 & 3000 & ?  & y & y & y & no & absorption \cr
HR 363 & 4590 & var & y & no & ?  & ? & wk em.\cr
\hline
Hen 18 & ? & ? & y & ?  & ?  & strong & ?\cr
Hen 134 & ? & ?  & ? & ? & ?          & moderate & ? \cr
ER Del & ?  & ?  & ?  & ?  & ?  & strong& ? \cr
\hline
Ref &1 &2 &3 &3 &4 &5&6 \cr
\hline

\end{tabular}
\vskip 0.5cm

{\small References:
(1) Jorissen et al., 1998, A\&A 332,
877; Griffin, 1984, The Observatory 104, 224; Carquillat et al., 1998, A\&AS 131,
49; Paper II (2) Jorissen et al., 1996, A\&A 306, 467
(3) Johnson et al., 1993, ApJ 402, 667 and
references therein; Ake, 1996, priv. comm.
(4) Ake et al., 1994. In:
A.W. Shafter (ed.) Interacting Binary Stars (ASP Conf. Ser. 56); Ake, 1996, priv. comm.
(5) Johnson \& Ake, 1989. In: H.R. Johnson,
B. Zuckerman (eds.) Evolution of Peculiar Red Giants (IAU Coll. 106)
Cambridge Univ. Press, p.371; Ake et al., 1991, ApJ 383, 842; this work
(6) Brown et al., 1990, AJ 99, 1930; 
Shcherbakhov \& Tuominen, 1992, A\&A 255, 215
}
\end{table*}

\begin{figure}[]
\resizebox{\hsize}{!}{\includegraphics[290,60][550,550]{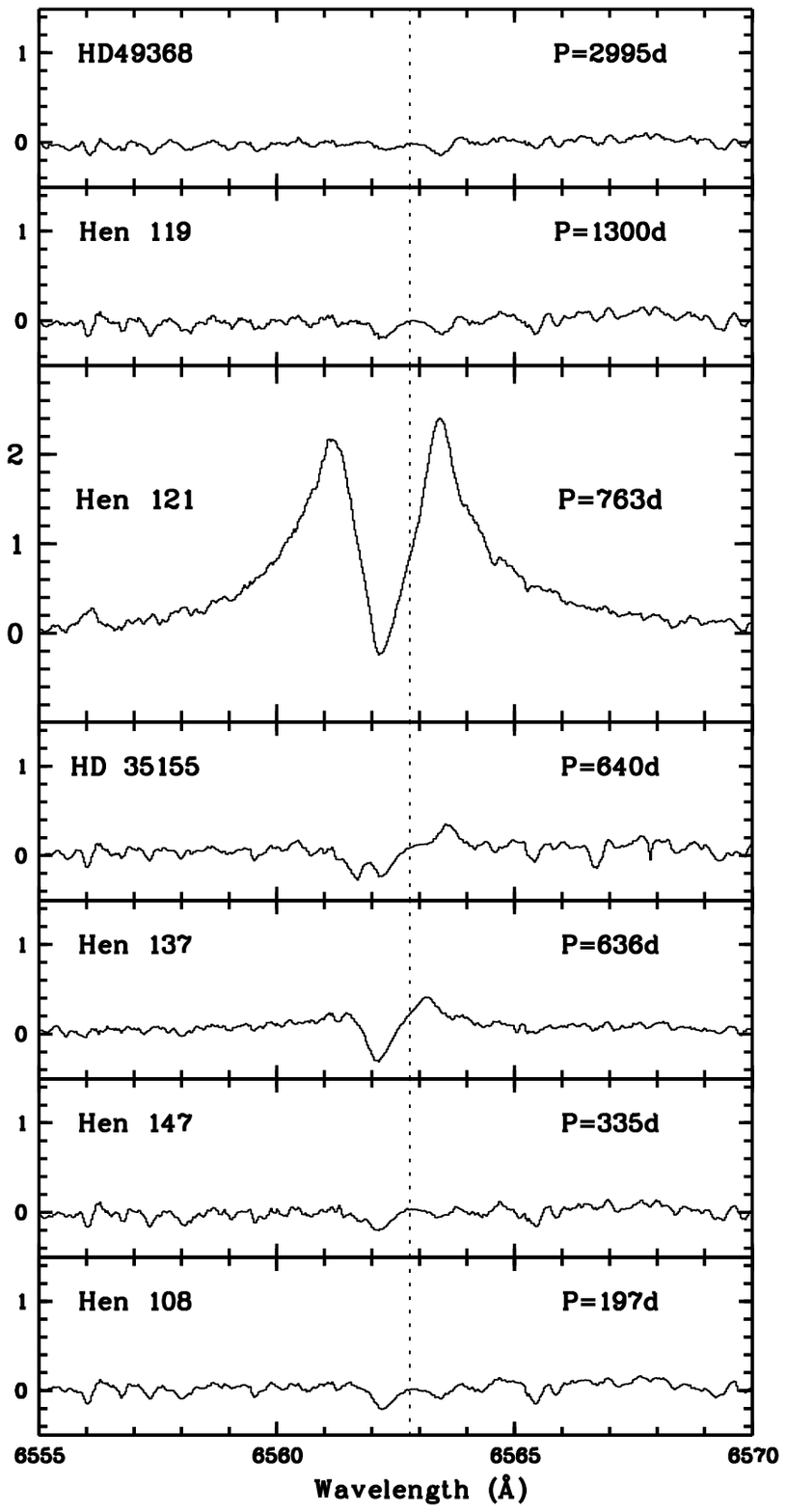}}
\caption[]{\label{Fig:HaP}
Same as Fig.~\protect\ref{Fig:HaHen6} for binary S stars, ordered from top to bottom 
according to decreasing  orbital periods (as indicated in the upper right corner). It is
clearly apparent that S stars with \Ha\ emission are found in the period range
600--1000~d
}
\end{figure}

\begin{figure}[]
\resizebox{\hsize}{!}{\includegraphics{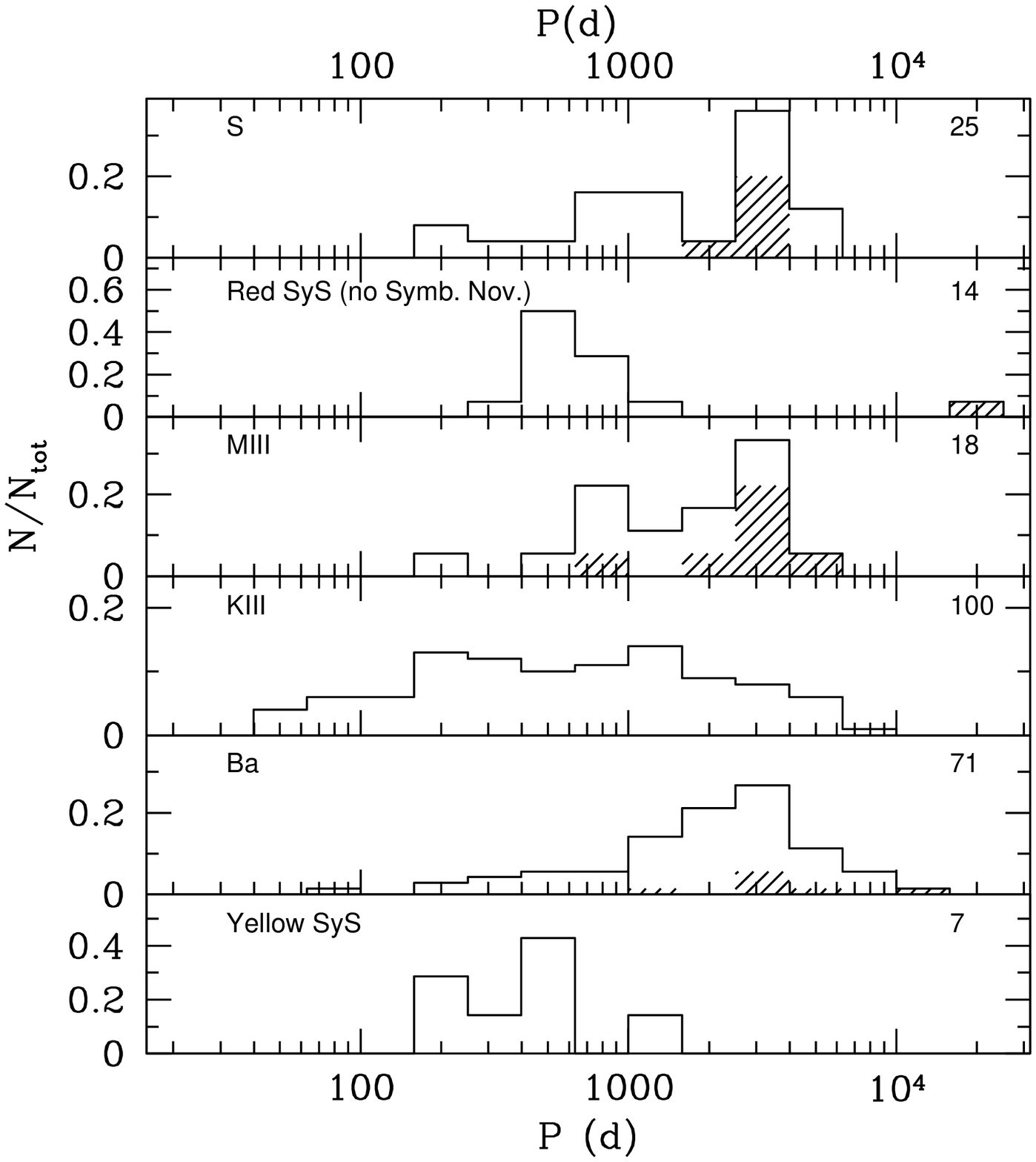}}
\caption[]{\label{Fig:PKM}
Comparison of the orbital period distributions for S stars \citep{Jorissen-VE-98}, red
symbiotics \citep[excluding symbiotic novae;][]{Muerset-99}, K giants
\citep{Mermilliod-96} and M giants (Jorissen et al., in preparation). 
The lower two
panels  present the orbital-period distribution for barium stars \citep{Jorissen-VE-98} and
yellow symbiotics \citep{Muerset-99}.
The shaded area marks stars with only a lower limit on their orbital period. The numbers in the upper right corner 
of each panel correspond to the sample size}
\end{figure}

\begin{figure}[]
\resizebox{15cm}{!}{\includegraphics{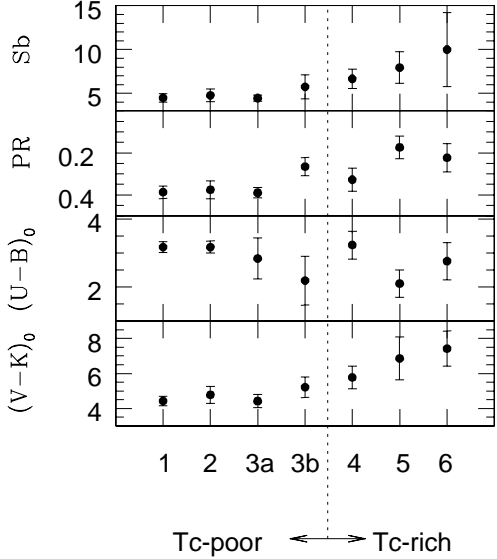}}
\vspace{-7cm}
\caption[]{\label{Fig:VK}
Comparison of luminosity and temperature indicators for binary S stars (groups 1,
2 and 3) and intrinsic S stars (groups 4, 5 and 6; see Paper III for a detailed description of
these groups). Symbiotic S stars (Hen 18, 121, 134, 137 and HD 35155) form 
group 3b, whereas binary S stars with no \protect\Ha\ emission (Hen 108, 119, 147 and
HD 49368) constitute group 3a. The color index $V-K$ is a good temperature indicator,
whereas the CORAVEL parameters $Sb$ (average spectral line width, in
\kms, corrected for the instrumental width) and $PR$ (average line depth) are good
luminosity indicators. 
The symbiotic S stars have temperature and luminosity indicators that are the most
extreme among binary S stars, and are close to the values characterizing intrinsic, more
evolved, S stars. On the contrary, binary S stars with no \protect\Ha\ emission (group
3a) are clearly much less evolved than the symbiotic S stars.  For comparison the
deredenned
$U-B$ color index is also shown
}
\end{figure}

The present data offer a unique opportunity to correlate the
symbiotic activity (diagnosed by the presence of \Ha\ emission) 
with the orbital period (Fig.~\ref{Fig:HaP}).
Neither the binaries with the shortest periods
nor those with the longest do exhibit any  \Ha\ emission.  Symbiotic activity seems
instead restricted to a rather narrow period range around 800~d ($\pm 200$~d;
see Table~\ref{Tab:activityP}). This conclusion is consistent with the 
distribution of orbital periods known so far among Henize S stars. There are 
8 orbital periods available (Table~1 of Paper~II), among which only 3
(Hen~18, 121, 183) fall in the symbiotic period range. Two of them are 
indeed symbiotics.
The third one (Hen~183) does not seem to show any
sign of symbiotic activity (normal $UBV$ color indices but no data
available on \Ha; Paper~II); its case is further discussed
below.
The other 5 systems have orbital periods
either shorter or larger than the symbiotic period range. Since binaries
with short periods are the easiest to discover, it is likely that
all short-period systems existing among the Henize
extrinsic S stars have been found. Thus, the small number of
symbiotics among the Henize sample just reflects their orbital-period distribution.   

The reason why symbiotic S stars seem restricted to the period range
600 -- 1000~d should now be understood, and in particular why systems
with shorter orbital periods are not symbiotics.
This question may actually be addressed by identifying the
parameters required to trigger symbiotic activity \citep[see also the
discussion in][]{Jorissen03}.
The key factor in this respect is the high luminosity of the
compact companion, that should exceed 10 \Lsun\
\citep[e.g.,][]{Yungelson-95}, in order to provide the high-energy photons
necessary to excite atoms in the red-giant wind. This
requirement in turn sets a constraint on the accretion rate by the compact star, which
should be large enough to deliver such a luminosity, either through nuclear burning at
the white dwarf surface or directly through accretion luminosity. 
Detailed hydrodynamical simulations of mass transfer in symbiotic binaries 
\citep{Theuns-96, Mastrodemos98}
have shown that the
Bondi-Hoyle accretion regime generally predicts accretion rates that are an order of magnitude too large. 
It provides however a convenient analytical formulation of the mass accretion rate that may be used 
to identify key parameters:
\parbox{7cm}{
\begin{eqnarray*}
\label{Eq:B-H}
\dot{M}_{acc, B-H}  &=&  -\dot{M}_{\rm wind} \; \frac{\eta}{A^2} \; \left( \frac{G\; M_2}{v_{\rm wind}^2} \right)^2 
 \nonumber\\
&&  \times~\frac{1}{\left[1 + (\frac{v_{\rm orb}}{v_{\rm wind}})^2 + \left(\frac{c}{v_{\rm wind}}\right)^2\right]^{3/2}} 
\end{eqnarray*}
}\hfill\parbox{1cm}{\begin{eqnarray}\end{eqnarray}}
where
$M_2$ is the mass of the accreting star, $\eta \sim 1$  if the Bondi-Hoyle accretion regime applies 
(i.e., $v_{\rm wind} /v_{\rm orb} >> 1$),  $\eta \sim  0.1$  otherwise (if $v_{\rm wind} /v_{\rm orb} \la 1$). 

Eq.~\ref{Eq:B-H} thus shows that three parameters are of prime importance in fixing the accretion rate: 
the wind velocity of the giant star $v_{\rm wind}$, which according to \citet{VanLoon-00} is proportional to
$L_{\rm 1}^{1/4}\;\sqrt{Z}$ ($Z$ is the metallicity and $L_{\rm 1}$ the luminosity of the giant), the orbital period
$P$ which is related to both the orbital separation $A$ and the  orbital velocity
$v_{\rm orb}$, and the wind mass loss rate $\dot{M}_{\rm wind}$.

The supposedly low mass-loss rates characterizing S stars in short-period
systems  probably  account  for their absence of symbiotic activity.
Indeed, only compact, i.e, warm and low-luminosity giants may be hosted 
by the shortest-period (200 -- 500~d) systems. More evolved giants are too big to hold
within their Roche lobe in short-period systems. 
This statement is confirmed by the empirical correlation between spectral types and
orbital periods observed among symbiotic systems by \cite{Muerset-99}, late spectral types being restricted to 
systems with the longest periods. It is also apparent from the comparison of the period distributions of K and M 
giants (Fig.~\ref{Fig:PKM}): no system with an orbital period shorter than about 400~d is observed among M giants,
whereas the lower period cutoff for K giants is only about 40~d.

Fig.~\ref{Fig:PKM} moreover 
reveals that the period range 600 -- 1000~d where most symbiotic S stars are found is also 
where most red symbiotic stars are located (when excluding symbiotic novae and symbiotic Miras). 
Those red symbiotic stars are in turn found in the short-period tail
of the distribution of M giants. Similarly, yellow symbiotic
stars occupy the short-period tail of the distribution of Ba stars (Fig.~\ref{Fig:PKM}). 
Symbiotic stars are thus found among
binaries having the  shortest allowed periods  given the evolutionary stage of the cool
component.
Binary S stars with the longest orbital periods do not exhibit \Ha\ emission (Fig.~\ref{Fig:HaP} and
Table~\ref{Tab:activityP}) since, for a given mass loss rate, the accretion rate is smaller in wider systems
(Eq.~\ref{Eq:B-H}). Table~\ref{Tab:activityP} reveals, however, that very long-period systems like
HR~363 and HD~49368 nevertheless exhibit other signatures of symbiotic activity (for instance, in
the form of X-rays or UV continuum), despite the absence of  \Ha\ emission.  This is not surprising since   
\Ha\  emission is not as sensitive a diagnostic of symbiotic activity as are the UV or X-ray emissions, as confirmed for
instance by the absence of \Ha\ emission at some epochs for the classical -- albeit rather weak -- symbiotic star EG~And.

Fig.~\ref{Fig:VK} compares  the properties of intrinsic S stars, symbiotic S stars and non-symbiotic extrinsic
S stars, and provides an indirect confirmation of the above qualitative conclusion that symbiotic S stars ought to
lose more mass than non-symbiotic, binary  S stars, and thus  presumably be somewhat more evolved. The
$V-K$ index is a good temperature indicator, and Fig.~\ref{Fig:VK} indicates that the symbiotic S stars (group 3b in
Fig.~\ref{Fig:VK}) are among the coolest extrinsic S stars. Observable diagnostics of luminosity include the
CORAVEL $Sb$ and $PR$ parameters, which measure the average line width and
depth, respectively (see Paper II for a more detailed description of these indicators). 
Fig.~\ref{Fig:VK} shows that the symbiotic S stars have values of  $Sb$ and $1-PR$
that are the largest among binary S stars, even close to those
characterizing the more evolved intrinsic S stars. 
Thus, both the temperature and luminosity indicators suggest that the
symbiotic S stars are the most evolved  among the binary S stars, and thus may indeed
be expected to suffer from the largest mass loss rates.
The above analysis also justifies why Hen~183 does not show any sign
of symbiotic activity, despite an orbital period of about 890~d. 
With $Sb=4.7$~\kms, $PR=0.37$ and $(V-K)_0=4.77$, Hen~183 appears somewhat less evolved than
symbiotic S stars (Fig.~\ref{Fig:VK}).

\section{Conclusions}

Two strong (Hen 18 and Hen 121) and two weak (Hen 134 and Hen 137)
symbiotic stars were found among the binary S stars from the Henize sample,
as revealed by their strong or moderate \Ha\ emission. Their \Ha\ profiles
are typical of symbiotic stars, more precisely of class S-3 as defined by
\citet{VanWinckel-93}. Hen 18 and Hen 121 are also characterized by very
blue $B-V$ and $U-B$ indices. The eclipse of the Balmer
  ultraviolet continuum observed for Hen~121 clearly indicates that
  the emission region is tied to the companion.  
The symbiotic S stars, when diagnosed by
their \Ha\ emission, are restricted to the period range 600 -- 1000~d, also
typical of red symbiotic stars (excluding symbiotic novae and symbiotic
Miras), and which represents the short-period tail of the distribution of
binary M giants. 
Moreover, there is compelling evidence that symbiotic S stars are
somewhat more evolved than non-symbiotic binary S stars.
These properties may be easily understood since the cool components in
symbiotic systems need to lose mass at a rate large enough to
power the high luminosity ($L > 10$~L$_\odot$) of the compact companion
and thereby to trigger the observed symbiotic activity.

\begin{acknowledgements}
We thank G. Burki and M. Burnet for securing the photometric observations,
M. K\"urster for help with the computation of line bisectors, and 
H.M. Schmid for enlightening discussions about the physics of symbiotic stars.
\end{acknowledgements}

\bibliographystyle{apj}
  
\bibliography{/home/bibtex/ajorisse_articles,/home/bibtex/svaneck_books}



\end{document}